# Towards Real-World RUL Prediction for Aircraft Compressors Under Variable Conditions with Physics-Guided Domain Adaptation

Yang Zhang[1], Shashvat Prakash[2], Jiong Tang[1,†]

1 School of Mechanical, Aerospace, and Manufacturing Engineering, University of Connecticut

2 RTX-Collins Aerospace, Avionics

## Abstract

Remaining useful life prediction for aircraft centrifugal air compressors in real commercial operations presents challenges that controlled benchmark datasets are not designed to expose. Sensor signals recorded during flight reflect a superposition of genuine component degradation and continuously varying operating conditions, and no single channel can be assumed a priori to carry a reliable degradation signature. Beyond signal quality, models trained on one subset of a fleet generalize poorly to unseen aircraft and installation positions, a cross-domain problem whose practical severity has received limited attention. This paper presents a framework that addresses both challenges through a sequential pipeline of physics-guided processing and adaptive temporal encoding. A population-level indicator selection procedure identifies surge margin as the most consistent degradation-relevant signal across the fleet. Operating regime filtering and windowed aggregation recover the underlying health trend from operational noise, and two physically motivated features encoding current health level and cumulative degradation rate enable cross-unit comparison without requiring absolute signal values. A sinusoidal positional encoding provides lifecycle context while avoiding data leakage, and a structured cross-domain evaluation identifies positional encoding period mismatch as the primary mechanism of performance degradation under fleet transfer. An adaptive recalibration scheme addresses this mechanism by estimating the lifecycle scale of each target unit from early operational observations alone, without future information or labeled target data. The approach yields consistent and substantial improvement in cross-domain prediction accuracy across transfer scenarios of increasing domain distance, with a Gaussian Process Regression model further providing calibrated uncertainty estimates that directly support risk-informed maintenance scheduling.

[†] Corresponding author

## 1. Introduction

Modern commercial aircraft fleets generate vast quantities of sensor data through flight operational quality assurance (FOQA) systems, recording hundreds of parameters at sub-second intervals throughout every flight. The ambition of prognostics and health management (PHM) is to transform this continuous stream of operational data into actionable maintenance intelligence, identifying early signatures of component degradation and estimating how much useful life a component retains before it must be replaced (Zio, 2022). For an airline operator, reliable remaining useful life (RUL) estimates translate directly into measurable operational value. Unnecessary premature removals are avoided, unscheduled maintenance events that strand aircraft are prevented, and maintenance resources can be allocated prospectively rather than reactively. These benefits are particularly substantial for high-value components with long and variable service lives, where the difference between evidence-based and calendar-based maintenance decisions can represent meaningful reductions in both cost and operational disruption.

The research community has made considerable progress on data-driven RUL prediction, driven in large part by the availability of standardized benchmark datasets. The Commercial Modular Aero-Propulsion System Simulation (C-MAPSS) turbofan engine dataset (Saxena et al., 2008) and the PRONOSTIA bearing dataset from the IEEE PHM 2012 Challenge (Nectoux et al., 2012) have together enabled the development and systematic comparison of a wide range of methodologies, including deep recurrent architectures and attention-based transformers (Zhang et al., 2023). Semi-supervised and hybrid deep frameworks have broadened applicability to settings where labeled run-to-failure data are scarce (Ellefsen et al., 2019). Hybrid frameworks that combine physics-based performance models with data-driven learning have been shown to extend RUL prediction horizons substantially compared to purely data-driven baselines, and to require less training data to achieve comparable accuracy (Arias Chao et al., 2022). More recently, graph neural networks (GNNs) have emerged as a prominent framework for RUL prediction by modeling the spatial dependencies among sensor channels as graph-structured data. These approaches construct adjacency matrices either from physical engine topology or from learned sensor correlations and apply graph convolution or attention mechanisms to aggregate inter-sensor information before temporal modeling (Wang et al., 2025a). Representative methods such as the adaptive dynamic fusion GNN (Cao et al., 2025), the GCN-based parallel route model (Song et al., 2024), the spatiotemporal graph attention tensor network (Liu et al., 2025a), and the operating condition-aware heterogeneous GNN (Tian et al., 2026) have demonstrated state-of-the-art performance on C-MAPSS and N-CMAPSS benchmarks. Transformer-based and state-space architectures have further advanced temporal feature extraction, with physics-informed transformer designs (Xiao et al., 2026) and cross-attention multi-scale state space models (Zhang et al., 2026) and physics-guided signal compression frameworks (Zhou et al., 2026) establishing new benchmarks on the same datasets. A comprehensive account of the current state of data-driven prognostics, including the open challenges of uncertainty quantification, robustness, and real-world feasibility, is provided by Salinas-Camus et al. (2025). Gaussian process regression has been applied to RUL prediction specifically for its ability to provide calibrated predictive uncertainty, a property that directly informs risk-based maintenance decisions (Xu et al., 2022; Benker et al., 2021; de Pater et al., 2022; Mitici et al., 2023). Progress on these benchmarks has been real and measurable. However, several structural characteristics of benchmark datasets differ in fundamental ways from the conditions encountered when deploying predictive maintenance on real in-service aircraft, and the gap between benchmark performance and genuine operational deployability remains substantial.

This gap has two principal dimensions. The first concerns signal quality and health indicator construction. Simulated datasets are generated by a known degradation model that determines the structure of the health signal, and any added noise is typically stationary and well-characterized. Real FOQA data, by contrast, reflects the full complexity of commercial operations. Every flight involves a distinct route, payload, weather encounter, and crew technique, all of which influence the sensor readings recorded onboard. Health signals extracted from real operational data therefore carry a substantially higher ratio of operational variability to degradation signal than their simulated counterparts (de Pater and Mitici, 2023; Xue et al., 2025). In this setting, constructing a reliable health indicator from multi-sensor observations is itself a non-trivial problem. No single sensor channel can be assumed a priori to carry a consistent degradation signature, and the relative informativeness of different channels varies across units within the same fleet (Wang et al., 2023; Liu et al., 2022). Although GNN-based methods model inter-sensor correlations and have shown promise in capturing multi-source degradation information, their graph representations are learned end-to-end from benchmark data and presuppose that degradation signatures are distributed across multiple sensor channels simultaneously. When applied to real operational data in which every channel is dominated by operating-condition variation rather than degradation, the learned graph structure reflects operational coupling rather than mechanical degradation, and there is no guarantee that the resulting representation satisfies the physical properties required of a reliable health indicator. Methods that fuse multiple sensor streams through attention or gating mechanisms have demonstrated improvements in indicator quality (Ni et al., 2022; Zhou and Tang, 2023), but these approaches rely on end-to-end learned representations and do not guarantee that the resulting indicators satisfy the physical properties expected of a genuine degradation measure. Trendability and monotonicity have been established as foundational evaluation criteria precisely because a candidate indicator must reflect progressive degradation rather than operational fluctuation to be meaningful for prognostics (Javed et al., 2014; Zhang et al., 2016). Recovering indicators with these properties from real operational data requires domain knowledge about the physical degradation mechanism, not only expressive model architectures. A related limitation applies to methods for time-varying operating conditions (Liu et al., 2025a; Tian et al., 2026; Liu et al., 2025b): these approaches address condition-induced distribution shift within the feature space, but they still assume that a coherent degradation signal can be extracted from multi-sensor data, a premise that does not hold when the signal-to-noise ratio of degradation relative to operations is as unfavorable as it is in real FOQA records.

The second dimension concerns cross-unit generalization. On benchmark datasets, components within the same operating condition subset tend to degrade at broadly similar rates. In a real fleet, the same component design may exhibit total service lives differing by a factor of three or four, reflecting differences in individual operating history, maintenance quality, and inherent manufacturing variation. A model trained on one subset of the fleet and applied to another must therefore generalize across a distribution of degradation behaviors that is substantially wider than what it was trained on. Domain adaptation methods for RUL prediction have been proposed to address distributional shift between source and target operating conditions by learning domain-invariant feature representations through adversarial training or statistical alignment (Fu et al., 2021; Ding et al., 2022). Methods based on domain-weighted distribution adaptation have extended this to time-varying condition scenarios where multiple operating regimes alternate during degradation (Liu et al., 2025b), and self-supervised pretraining has been shown to enable transfer of diagnostic knowledge across aircraft types with limited labeled target data (Wang et al., 2025b). Incorporating the operational profile of each fleet segment into the alignment process has been shown to improve transfer accuracy when mission-level characteristics differ between source and target (Nejjar et al., 2024). A particularly practical setting arises when only early degradation observations are available from the target unit, and transfer must be achieved without access to the target failure trajectory (Li et al., 2024). However, most existing methods address distribution shift in feature space and leave unexamined a structurally different form of mismatch that arises specifically in fleet prognostics. When a model encodes lifecycle position through a fixed temporal representation calibrated to the lifespan distribution of the source fleet, applying it to target units with substantially different total service lives introduces a systematic prediction error that is not reducible by feature alignment alone. This mechanism, which we identify as lifecycle-scale mismatch, calls for a recalibration strategy that operates at the level of the temporal encoding rather than the feature distribution.

The centrifugal air compressor (CAC) of commercial aircraft presents a particularly instructive case study for both challenges. The CAC supplies pressurized bleed air for cabin conditioning, anti-icing, and other pneumatic loads, extracting airflow from the engine compressor stage. Its primary health indicator, surge margin, measures the aerodynamic distance between the current operating point and the surge line, the boundary beyond which flow instability and potentially damaging surge events occur. As the compressor degrades through blade erosion, fouling, and clearance growth, the available surge margin decreases progressively over thousands of flight hours, and this decline is both real and measurable. However, surge margin also varies strongly with the instantaneous operating point of the compressor, which changes continuously with flight phase and throttle setting. The resulting sensor signal is noisy in a way that is physically meaningful but analytically inconvenient. The noise is not random but structured, reflecting the operational envelope of the aircraft, and no simple filter removes it without simultaneously attenuating the degradation information it contains. Embedding physical knowledge of the degradation mechanism into the feature engineering pipeline, rather than relying on learned representations alone, has been shown to produce more interpretable and generalizable degradation models in analogous aerospace systems (Ma et al., 2023). The importance of grounding health indicator construction in physical domain knowledge is further underscored by work on real aero-engine flight data, where operating regime filtering has been shown to be a prerequisite for recovering coherent degradation signals from otherwise noisy FOQA records (Xiao et al., 2026). Real run-to-failure data from in-service aviation components of this kind is rarely available in the published PHM literature, owing to the proprietary nature of fleet data and the long component lifespans involved, making the physics of CAC degradation a natural guide for methodological choices that cannot be fully validated by data volume alone.

This paper presents a framework developed and evaluated on real-world run-to-failure compressor records from real commercial operations, spanning different aircraft and installation positions. This dataset represents a resource that is uncommon in the published PHM literature, and its structure naturally supports a study of cross-domain generalization within a genuine operational context. The proposed framework addresses the two challenges described above through a sequential methodology. A population-level indicator selection procedure identifies surge margin as the most consistent degradation-relevant signal across the fleet. Operating regime filtering and windowed aggregation recover the underlying health trend from operational noise, and two physically motivated features encode current health level and cumulative degradation rate in a form that enables cross-unit comparison without relying on absolute signal values. A sinusoidal positional encoding provides lifecycle context while avoiding data leakage from future observations. A structured cross-domain evaluation then identifies lifecycle-scale mismatch as the primary mechanism of performance degradation under fleet transfer, and an adaptive recalibration scheme addresses this mechanism by estimating the lifecycle scale of each target unit from its early operational observations alone, without requiring future information or labeled target data. Prediction uncertainty is quantified through Gaussian process regression, providing calibrated confidence intervals that directly support risk-informed maintenance scheduling.

The remainder of the paper is organized as follows. Section 2 examines the properties of real-world compressor degradation data, using the dataset itself to motivate the methodological choices made throughout. Section 3 presents the full methodology, from signal conditioning through feature engineering to the adaptive positional encoding scheme. Section 4 reports experimental results including a feature ablation study, a structured cross-domain evaluation, and a comparison across multiple prediction models. Section 5 concludes with a discussion of practical implications and directions for future work.

## 2. Aircraft Compressor Fleet Data and the Challenges of Real-World Prognostics

This section examines the characteristics of real in-service compressor records that render RUL prediction challenging. Three challenges are identified. The first is the absence of a known health indicator: no sensor channel can be assumed a priori to carry a reliable degradation signature. The second is the dependence of candidate indicators on the instantaneous operating regime, which makes any naive use of raw measurements unreliable without physical conditioning. The third is the systematic domain shift in degradation behavior across aircraft and installation positions, which any fleet-deployed model must overcome.

### 2.1 Data description and operational context

The centrifugal air compressor occupies a critical position in the pneumatic architecture of commercial aircraft, supplying pressurized bleed air for cabin conditioning, anti-icing systems, and other essential loads throughout every phase of flight. A single unscheduled removal, triggered for instance by an in-flight cabin pressurization warning, carries consequences that extend well beyond the cost of a workshop visit, disrupting flight schedules, requiring immediate crew reassignment, and imposing passenger compensation obligations. The economic and operational incentive to predict compressor end of life accurately is therefore substantial, and it motivates the development of a prognostic framework grounded in real operational data rather than controlled laboratory conditions.

The present study uses run-to-failure compressor records collected through the Collins Aerospace Ascentia health management system, which records flight operational quality assurance (FOQA) parameters at sub-second intervals throughout every flight. Each record accumulates measurements continuously from the compressor's entry into service through its final removal, capturing the full degradation trajectory under real commercial operating conditions. The dataset spans multiple aircraft and compressor installation positions, providing a natural structure for studying how predictive models behave when deployed across a diverse fleet rather than on a single homogeneous population. The breadth of this dataset is itself unusual in the published prognostics literature, where run-to-failure records from in-service aviation components are rarely available due to the proprietary nature of fleet data and the long service lives involved. Figure 1 provides a schematic illustration of the aircraft pneumatic system and compressor installation configuration used in this study. Due to commercial confidentiality obligations with the data provider, certain details including aircraft type, tail numbers, fleet size, and exact installation positions have been withheld. The anonymized data is sufficient to support the analyses presented herein.

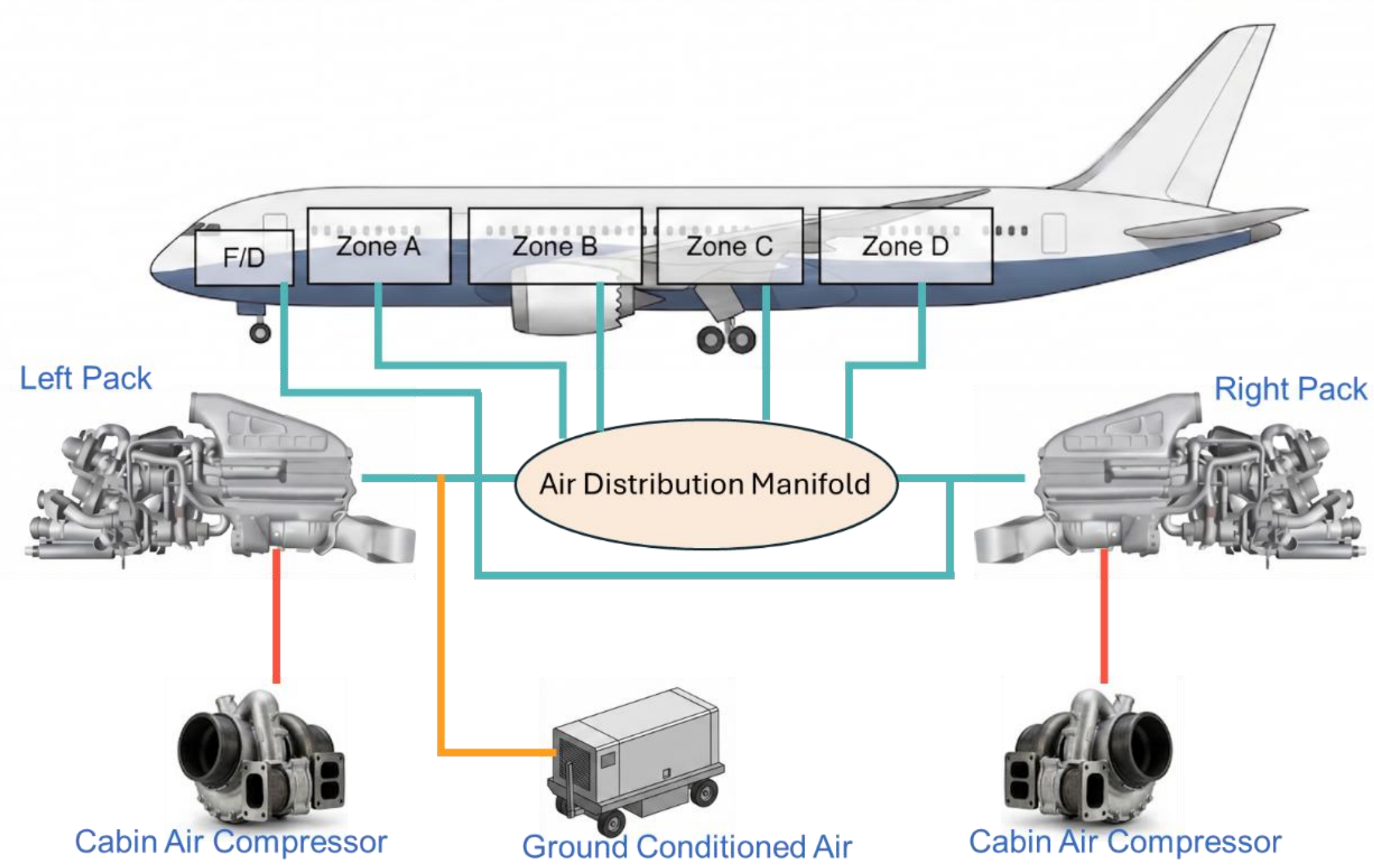

Figure 1. Aircraft pneumatic system schematic showing compressor installation positions and bleed air distribution architecture.

The raw sensor records for a representative compressor, shown in Figure 2, span all available channels including rotational speed, inlet and outlet pressures and temperatures, mass flow rates, power inputs, and derived aerodynamic quantities, etc. The dominant visual impression across every channel is one of extreme variability. Measurements swing across wide ranges from one flight segment to the next, and no channel exhibits a visually obvious monotonic trend over the full-service life. This variability is not noise in the statistical sense. It is physically meaningful variation driven by the real diversity of operating conditions encountered in commercial service, where different routes, altitudes, payloads, and ambient temperatures impose different aerodynamic demands on the compressor within every flight and across the fleet. The raw data therefore presents an immediate and fundamental obstacle to health monitoring, because any signal that responds to operating conditions will appear to change substantially even when the mechanical condition of the compressor is entirely unchanged.

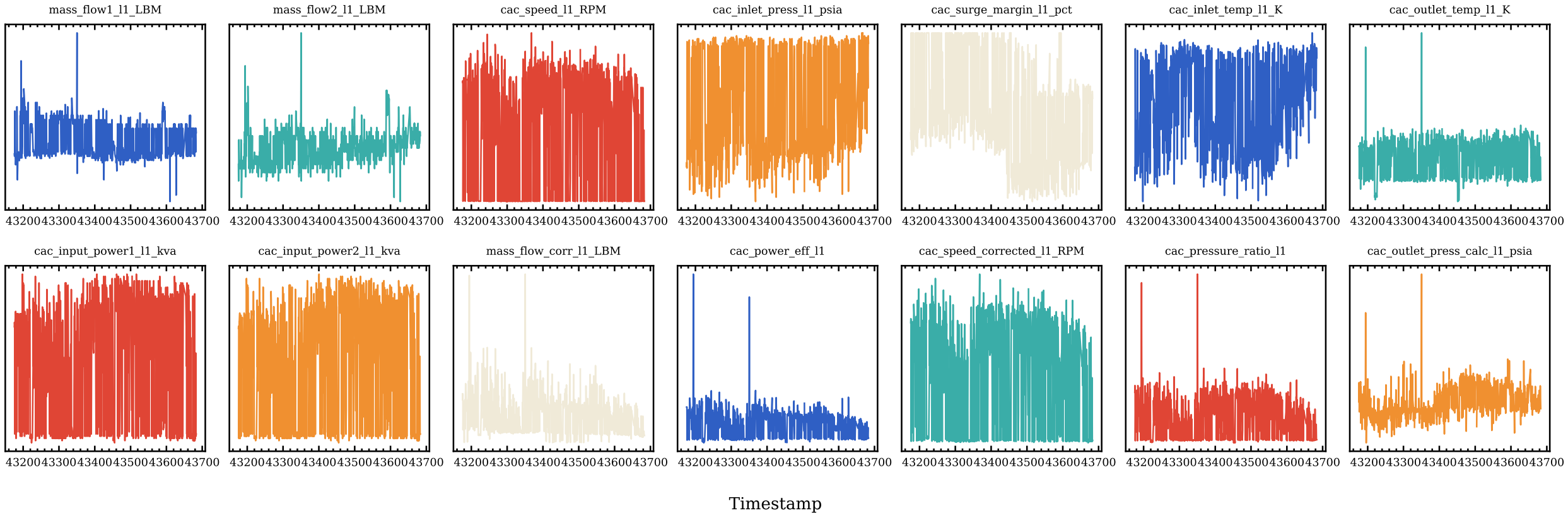

Figure 2. Raw FOQA sensor recordings for a representative compressor across its full-service life.

### 2.2 Signal identification and conditioning under operational variability

The first challenge confronting any prognostic system in this setting is that of identifying which sensor channels carry genuine degradation information when all channels are simultaneously influenced by operating conditions. A real aircraft compressor generates measurements across many channels at once, and there is no guarantee a priori that any of them carries a reliable degradation signature. Each channel reflects a superposition of the compressor's mechanical health state and the instantaneous operating conditions at the moment of measurement. A reading may be large or small because the compressor is healthy or degraded, or simply because the aircraft is climbing rather than cruising, or because the ambient temperature at altitude happens to be unusually low. Distinguishing these contributions by inspecting a single compressor record is not possible, because any apparent trend on one unit may reflect the particulars of that unit's operational history rather than a genuine degradation signature shared across the fleet. A valid identification procedure is necessary to therefore operate at the population level, evaluating signal behavior consistently across the full diversity of units and service histories.

The difficulty of this identification problem is compounded by the variable operating conditions under which commercial aircraft compressors operate throughout their service lives. Any sensor channel on a compressor reflects not only the mechanical health of the component but also the instantaneous aerodynamic and thermodynamic state imposed by the current flight phase, altitude, ambient temperature, and throttle setting. As these conditions change continuously across every flight and differ systematically across routes and missions, the resulting sensor signals carry a superposition of genuine health information and large-amplitude operational variability that cannot be separated by inspecting individual measurements. To make this concrete, consider surge margin, which measures the normalized distance between the operating point and the surge boundary on the compressor characteristic map. As a physical quantity it is directly responsive to blade erosion, fouling, and tip clearance growth, all of which shift the surge boundary adversely and reduce the available margin over the service life. At the same time, surge margin is strongly sensitive to corrected speed. During climb and cruise at high corrected speeds the operating point approaches the surge boundary and values are characteristically low, while during descent and approach at low corrected speeds the operating point retreats and values are high. The within-regime variability induced by these operating condition changes is comparable in magnitude to the total degradation-induced decline over thousands of flight hours, so that individual measurements carry almost no information about health state without conditioning on the concurrent operating regime. This situation is qualitatively different from what standardized benchmark datasets present, where degradation-induced changes dominate measurement variability and individual observations are directly informative. In a real operational setting, any candidate health indicator will exhibit this same entanglement between degradation and operating condition, and a population-level identification procedure must be capable of evaluating signal behavior across the full diversity of operating histories before any channel can be trusted to carry consistent degradation-relevant information.

The prognostic framework must therefore address two coupled subproblems before any regression model can be meaningfully applied. It must first identify, from among the available sensor channels and in a way that is valid across the full fleet population, which signals carry consistent degradation-relevant information. It must then extract the underlying health trend from those signals, separating the slowly evolving degradation component from the large-amplitude operational variability superimposed on it. Neither subproblem is encountered in benchmark prognostic studies, where the health indicator is determined by the experimental design and its degradation signal dominates measurement variability. Both must be solved here before the prediction problem can even be posed.

### 2.3 Cross-domain generalization across a heterogeneous fleet

A structurally distinct second challenge arises from the way degradation behavior varies systematically across a real fleet. Compressors of nominally identical design can exhibit substantially different degradation trajectories depending on their installation context. Units occupying different positions on the same aircraft operate at systematically different aerodynamic working points, because the bleed air demand profile, duct geometry, and pneumatic load distribution differ across positions. Across different aircraft, variation in route profiles, utilization intensity, and operational environments further compounds this heterogeneity. The consequence is that the statistical relationship between any health indicator and remaining useful life is not a fixed property of the component design. A model trained on one subset of the fleet learns a mapping from health features to residual life that is conditioned on the operating context of those particular units, and when deployed on a unit with a systematically different aerodynamic baseline, that mapping may no longer describe the observed behavior.

This constitutes a domain shift in the machine learning sense, where the joint distribution of features and labels in the target population differs structurally from that of the training population, rather than simply being noisier or more variable. Domain shifts of this character are largely absent from the benchmark datasets that have driven methodological development in prognostics over the past two decades, because those datasets test units under controlled and repeatable conditions designed to minimize such variation. In real fleet operations, domain shift is unavoidable. No airline accumulates complete run-to-failure histories uniformly across all aircraft types and installation positions before deploying a predictive maintenance system, and any practically useful system must generalize from the units it has observed to those it has not.

The cross-domain challenge is further complicated by the constraints of the prognostic setting itself. Standard domain adaptation methods that rely on labeled target data or access to the full target lifecycle cannot be applied, because the target unit's failure has not yet occurred and its remaining life is precisely what the system is trying to predict. The information available from a target unit at any point during service is limited to its operational history up to that moment, and any recalibration of the model must be grounded in that early-phase data alone. How domain shift degrades prediction performance, what mechanisms drive that degradation, and whether early operational observations from the target unit contain sufficient information to close the performance gap are therefore the central empirical questions motivating the second component of the framework developed in this paper.

## 3. Proposed approach for predicting RUL of CAC system on aircraft

The framework developed in this paper addresses the challenges identified in Section 2 through a sequential pipeline of three stages, each motivated by a specific failure mode of simpler approaches. Raw FOQA sensor data first passes through a population-level indicator selection procedure that identifies which signals carry reliable degradation information across the fleet. The selected indicator is then conditioned on operating regime and aggregated over time to recover the underlying health trend, from which a pair of physically motivated features is constructed to enable cross-unit comparison. A positional encoding scheme augments these features with lifecycle context in a form that does not introduce data leakage. Finally, an adaptive recalibration procedure extends the encoding to compressors whose lifecycle scale differs from the training population, addressing the cross-domain challenge without requiring any labeled data from the target unit. The resulting feature vectors are passed to regression models that produce calibrated RUL estimates with associated uncertainty. Figure 3 shows the overall pipeline.

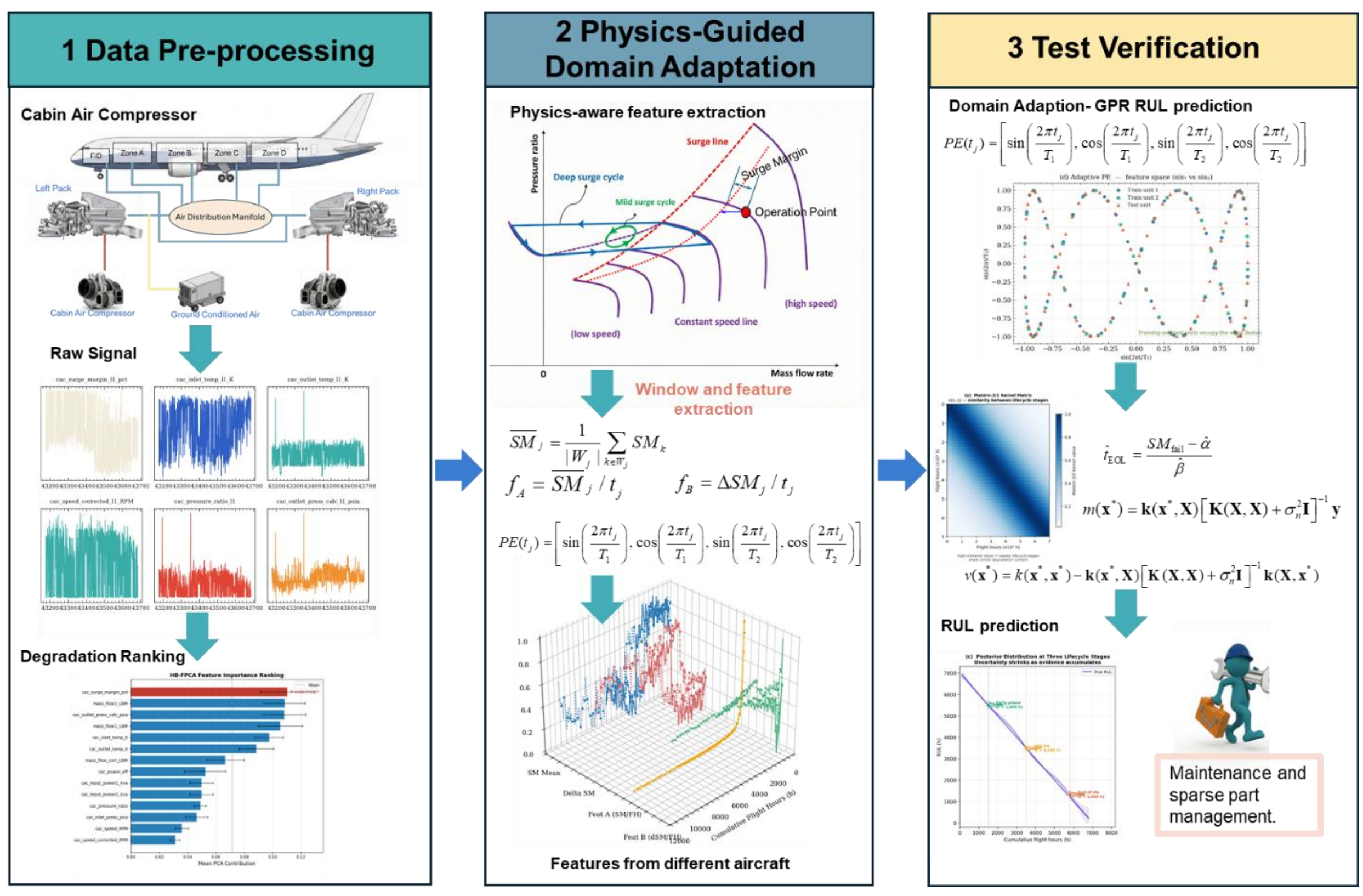

Figure 3. Overall methodology pipeline from raw FOQA data to calibrated RUL prediction.

### 3.1 Population-level health indicator selection

The starting point of the framework is the question of which sensor channel to use as a health indicator. As established in Section 2.2, inspecting a single compressor record is insufficient to answer this question reliably, because any apparent trend on one unit may reflect the particulars of that unit's operational history rather than a genuine degradation signature. What is needed is a method that evaluates each channel's degradation relevance consistently across the full population of service events, simultaneously accounting for how much variance the channel carries, how monotonically it evolves over service life, and how closely its variation tracks the accumulation of flight hours rather than instantaneous operating condition changes.

To this end, a Degradation-Aware Hybrid Basis Function Principal Component Analysis (DA-HB-FPCA) framework is proposed. For each service event $k$ containing $N_k$ observations, the $p$ available sensor channels are first standardized to zero mean and unit variance within the event to remove sensor-level scaling differences. The standardized data matrix $\mathbf{X}^{(k)} \in \mathbb{R}^{N_k \times p}$ is constructed as

$$\tilde{x}_{ij}^{(k)} = \frac{x_{ij}^{(k)} - \mu_j^{(k)}}{\sigma_j^{(k)}} \tag{1}$$

where $\mu_j^{(k)}$ and $\sigma_j^{(k)}$ are the sample mean and standard deviation of channel $j$ in event $k$. The sample covariance matrix is then

$$\mathbf{\Sigma}^{(k)} = \frac{1}{N_k - 1}\left(\tilde{\mathbf{X}}^{(k)}\right)^{\mathrm{T}} \tilde{\mathbf{X}}^{(k)} \tag{2}$$

Eigen-decomposition of $\mathbf{\Sigma}^{(k)}$ yields eigenvectors $\mathbf{u}_1^{(k)},\ldots,\mathbf{u}_p^{(k)}$ with corresponding eigenvalues $\lambda_1^{(k)} \geq \cdots \geq \lambda_p^{(k)}$. The number of retained components $m^{(k)}$ is determined by the minimum number required to explain a fraction $\tau$ of total variance,

$$m^{(k)} = \min\left\{ m : \frac{\sum_{i=1}^{m} \lambda_i^{(k)}}{\sum_{i=1}^{p} \lambda_i^{(k)}} \geq \tau \right\} \tag{3}$$

The PCA contribution of channel $j$, which measures how much of the retained variance that channel explains, is

$$\tilde{c}_j^{(k)} = \frac{\sum_{i=1}^{m^{(k)}} \left|u_{ij}^{(k)}\right|}{\sum_{l=1}^{p} \sum_{i=1}^{m^{(k)}} \left|u_{il}^{(k)}\right|} \tag{4}$$

PCA contribution alone, however, rewards signals that are highly variable regardless of whether that variability is related to degradation. Two complementary degradation-specific criteria are therefore computed for each channel within each service event. The monotonicity score quantifies the degree to which the signal changes consistently in one direction over time,

$$\mathrm{Mon}_j^{(k)} = \frac{\left| \sum_{t=1}^{N_k - 1} \mathbf{1}(\Delta y_{jt} > 0) - \mathbf{1}(\Delta y_{jt} < 0) \right|}{N_k - 1} \tag{5}$$

where $\Delta y_{jt} = y_{j,t+1} - y_{j,t}$. The trendability score is the absolute Spearman rank correlation $\rho_s$ between the channel $x$ and cumulative flight hours $t$,

$$\mathrm{Trend}_j^{(k)} = \left| \rho_s\left(x_j^{(k)}, t^{(k)}\right) \right| \tag{6}$$

Each of the three scores is averaged across all $K$ service events to yield population-level estimates $\tilde{c}_j$, $\overline{\mathrm{Mon}_j}$, and $\overline{\mathrm{Trend}_j}$, which are then normalized to the unit interval via min-max scaling. The final degradation-aware score for each channel is a weighted combination of the three normalized criteria,

$$S_j = w_1 \tilde{c}_j + w_2 \overline{\mathrm{Mon}_j} + w_3 \overline{\mathrm{Trend}_j} \tag{7}$$

where $w_1 = 0.25$, $w_2 = 0.25$, and $w_3 = 0.50$. The trendability criterion is assigned the greatest weight because it most directly measures the temporal alignment between a signal and the progression of degradation, which is the defining property of a useful health indicator. The PCA contribution and monotonicity criteria serve as complementary filters that reward signals which are informationally rich and directionally consistent, while penalizing those whose variation is dominated by operating condition responses.

The results across all service events are shown in Figure 4. Surge margin achieves the highest population-level score by a clear margin, with its advantage attributable primarily to its consistently high trendability across the fleet. Several other channels, including mass flow rate and inlet temperature, achieve competitive PCA contributions, but none approaches surge margin in the trendability dimension, confirming that those channels carry more operating condition information than health state information. The selection of surge margin as the primary health indicator is physically consistent with established compressor aerodynamics: as blade erosion, fouling, and tip clearance growth accumulate over service life, the surge boundary shifts adversely, and the available surge margin progressively decreases irrespective of the specific flight condition being flown. The DA-HB-FPCA result therefore confirms what compressor physics predicts, and surge margin is adopted as the health indicator for all subsequent stages of the framework.

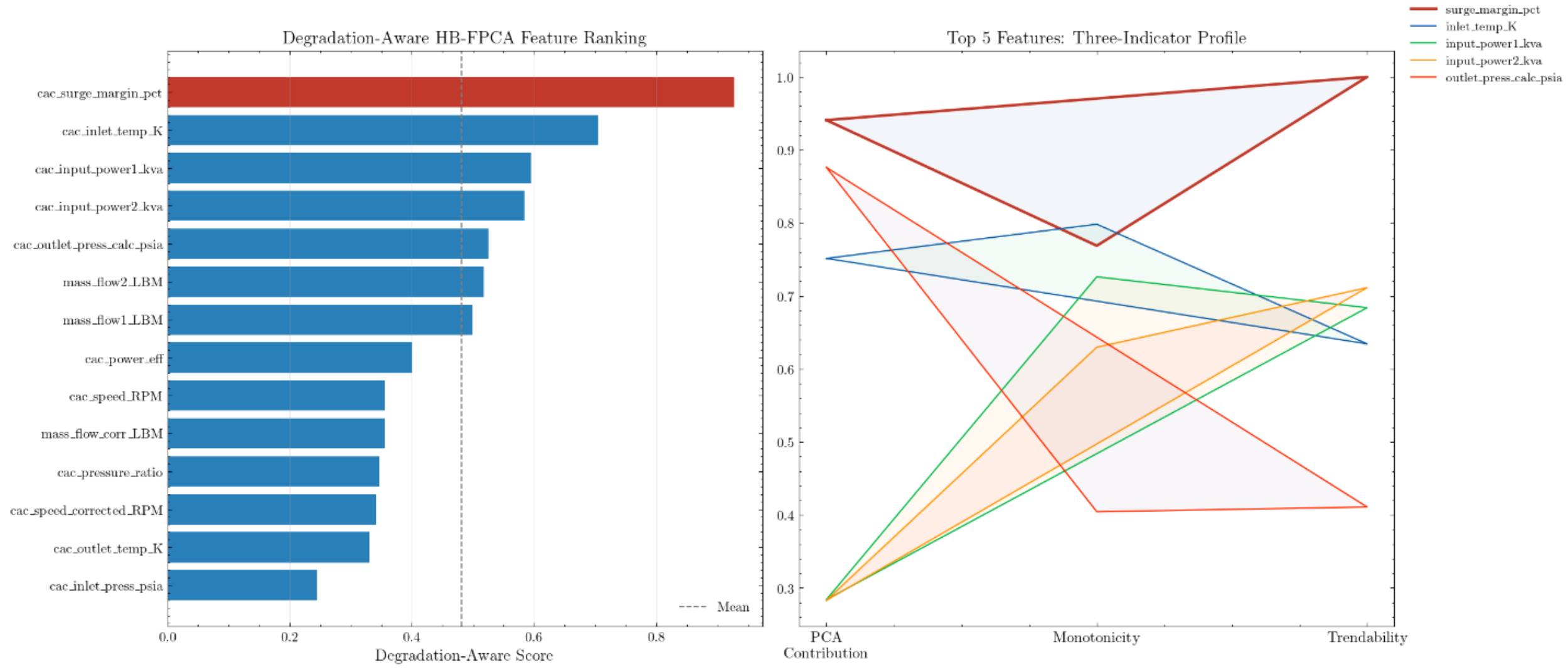

Figure 4. Population-level degradation-aware scores for all sensor channels.

### 3.2 Signal conditioning and physics-motivated feature engineering

Having identified surge margin as the primary health indicator, the next challenge is that this signal cannot be used directly for regression. As described in Section 2.2, surge margin at any given instant reflects both the health state of the compressor and the instantaneous aerodynamic operating regime, and the variability introduced by operating conditions is comparable in magnitude to the total degradation-induced change over the service life. Two processing steps are therefore required to recover a usable health trend: operating regime conditioning to remove the regime-induced variability, and temporal aggregation to suppress the residual flight-to-flight noise that remains within a given regime. The high-speed operating regime, corresponding to climb and cruise conditions, is selected for conditioning. As illustrated in Figure 5, at high corrected speeds the aerodynamic operating point sits closest to the surge boundary, maximizing the sensitivity of surge margin to changes in compressor health. At lower speeds the operating point retreats well away from the boundary, reducing this sensitivity and making measurements less informative about the degradation state. Restricting analysis to high-speed measurements retains the portion of the data where surge margin is most discriminative while discarding the portion dominated by regime-induced variability.

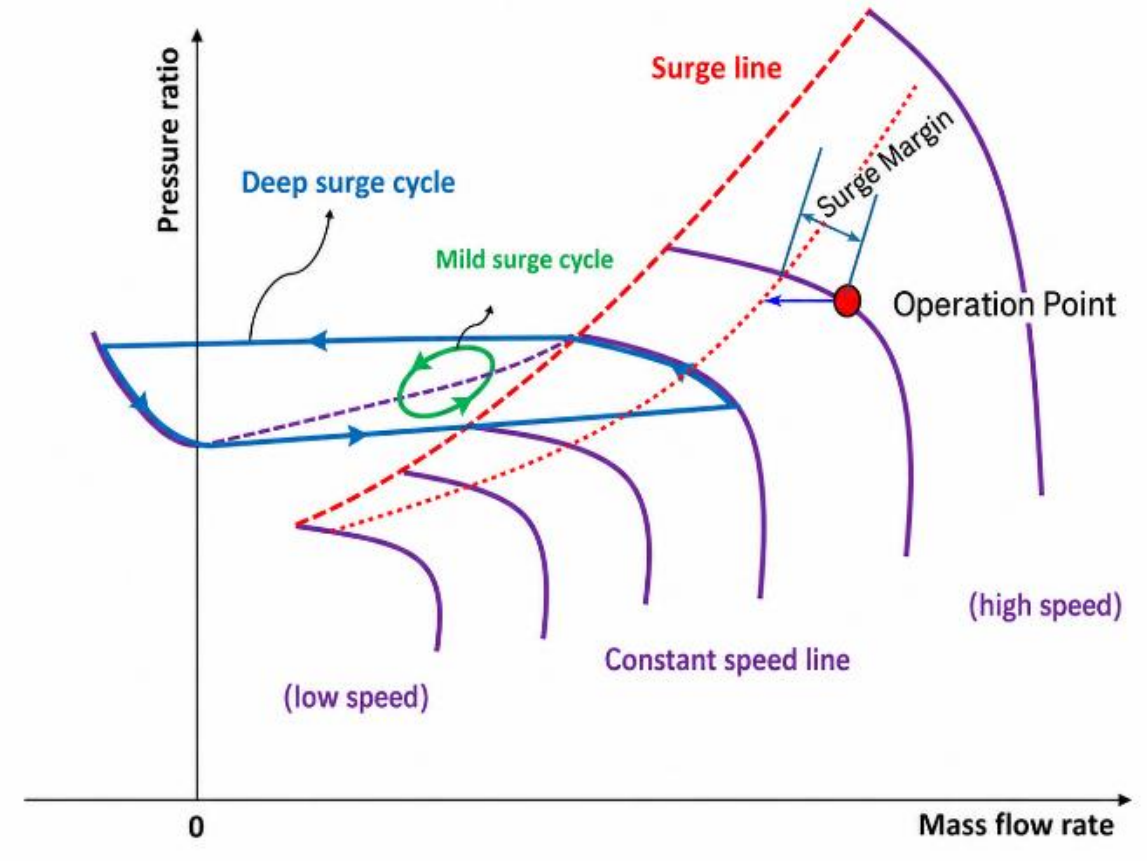

Figure 5. Physical illustration for surge margin.

Even within the high-speed regime, residual variability from differences in throttle setting, altitude, and ambient temperature persists across flights. A sliding window aggregation in the flight-hour domain suppresses this residual noise by averaging over many observations at comparable aerodynamic conditions. For each window $j$ ( $j = 1,...,N$ ) centered at flight hour $t_j$, the windowed mean surge margin is

$$\overline{SM}_j = \frac{1}{|W_j|}\sum_{k \in W_j} SM_k \tag{8}$$

where, $|W_j|$ is window size, $SM_k$ is the surge margin measurement at the $k$th data point within window $W_j$, and the summation runs over all high-speed measurements falling within the window and windows with fewer than a minimum number of valid observations are discarded. The resulting sequence $\{\overline{SM}_j\}$ constitutes a low-noise health indicator series that tracks the slowly evolving degradation trend while suppressing flight-to-flight operational variability.

Training a regression model directly on the windowed surge margin values is insufficient, however, because the absolute level of surge margin is not comparable across compressor units. Units installed at different positions on the aircraft operate at systematically different aerodynamic working points, resulting in different baseline surge margin levels at the start of each maintenance phase. As a consequence, the same absolute surge margin value can correspond to a unit near the beginning of its life or near the end, depending on its installation context. A model that processes absolute values will conflate units at different degradation stages and generalize poorly across positions.

Two physics-motivated features are constructed to address this ambiguity. Let $SM_0$ denote the surge margin measured in the first operational window, which is observable from the beginning of service without any future information. The cumulative degradation relative to the individual unit's own baseline is

$$\Delta SM_j = SM_0 - \overline{SM}_j \tag{9}$$

From this, two features are derived:

$$f_A = \overline{SM}_j / t_j \tag{10}$$

$$f_B = \Delta SM_j / t_j \tag{11}$$

$f_A$ encodes the current health level normalized by accumulated exposure. $f_B$ encodes the average degradation rate per flight hour over the observed service life. Because both features are referenced to the individual unit's own initial state and normalized by cumulative flight hours, they remain comparable across units with different baseline surge margins and different lifecycle stages, directly addressing the cross-unit ambiguity identified above.

### 3.3 Positional encoding for lifecycle context

The features $f_A$ and $f_B$ together characterize the health trajectory of a compressor up to the current observation, but they carry no explicit information about where in the overall lifecycle that observation falls. Two compressors with identical $f_A$ and $f_B$ values may be at very different lifecycle stages if one has degraded slowly from an early phase while the other has degraded more rapidly from a later one. Accurate RUL prediction requires the model to distinguish these cases, which demands a representation of temporal position within the lifecycle.

The most natural candidate, raw cumulative flight time $t_j$, introduces a data leakage problem. Normalized RUL is defined as

$$\mathrm{RUL}(t_j) = 1 - \frac{t_j}{t_{\mathrm{EOL}}} \tag{12}$$

so that flight time and RUL are linearly related through the total lifespan $t_{\mathrm{EOL}}$. If $t_{\mathrm{EOL}}$ is approximately known from the training distribution, a model with access to $t_j$ can recover RUL almost directly without engaging with the degradation signal at all. The apparent accuracy on in-distribution units would then reflect this arithmetic shortcut rather than any genuine understanding of degradation physics, and the model would fail on units whose lifespans differ from those seen during training.

To provide temporal context without this linear dependence, a sinusoidal positional encoding is adopted, drawing on the mechanism used in Transformer architectures. The current flight time is mapped to four bounded features at two frequencies,

$$PE(t_j) = \left[ \sin\left(\frac{2\pi t_j}{T_1}\right), \cos\left(\frac{2\pi t_j}{T_1}\right), \sin\left(\frac{2\pi t_j}{T_2}\right), \cos\left(\frac{2\pi t_j}{T_2}\right) \right] \tag{13}$$

with period parameters $T_1$ and $T_2$. The sinusoidal transformation preserves ordering information while removing the direct linear dependence on RUL, and the bounded output range ensures numerical stability across units with different total lifespans. The two frequencies provide temporal context at both the full-lifecycle scale and a sub-lifecycle scale, allowing the model to distinguish early, mid-life, and late-stage observations. The complete feature vector for window $j$ is

$$\mathbf{x}_j = \left[ f_{A,j}, f_{B,j}, \sin\left(\frac{2\pi t_j}{T_1}\right), \cos\left(\frac{2\pi t_j}{T_1}\right), \sin\left(\frac{2\pi t_j}{T_2}\right), \cos\left(\frac{2\pi t_j}{T_2}\right) \right] \tag{14}$$

All features are z-score normalized using statistics computed from the training set exclusively, and the same normalization parameters are applied to the test set without refitting.

### 3.4 Adaptive positional encoding for cross-domain transfer

The period parameters $T_1$ and $T_2$ define the lifecycle scale against which temporal position is encoded. When a model trained on one group of compressors is applied to a unit whose total lifespan differs substantially, the target unit traverses a different fraction of the encoding period over its life than the training units did. The sinusoidal features of the target unit therefore occupy regions of the encoding space that the model has not observed during training, and the temporal position information embedded in the encoding is effectively lost. This mismatch is structurally distinct from the operating condition variability and initial surge margin differences addressed earlier; it cannot be resolved by normalizing the health signal because it arises from the encoding of flight time itself.

The solution is to calibrate the encoding periods to the estimated lifespan of each individual unit rather than using a single fixed value for the entire fleet. For a target compressor, the total lifespan is estimated from its early operational observations without requiring any future information. A linear trend is fitted to the windowed surge margin values from the early phase of service

$$SM(t) = \alpha + \beta t, \quad t \in \text{early phase} \tag{15}$$

where slope $\beta$ represents the average rate of surge margin decline per flight hour during the early phase and $\alpha$ is the intercept of the fitted trend at the start of the observation window. The estimated end of life ($\hat{t}_{\text{EOL}}$) is the flight hour at which this trend reaches a terminal threshold $SM_{\text{fail}}$

$$\hat{t}_{\text{EOL}} = \frac{SM_{\text{fail}} - \hat{\alpha}}{\hat{\beta}} \tag{16}$$

where, $\hat{\alpha}$ and $\hat{\beta}$ are the estimated intercept and slope obtained by fitting Eq. (15) to the early-phase observations of the target unit. The terminal threshold is not assumed known a priori. It is learned from the training compressors as the mean ratio of the terminal surge margin to the initial surge margin at the time of removal

$$r_{\text{fail}} = \frac{1}{N} \sum_{i=1}^{N} \frac{SM_{\text{terminal},i}}{SM_{0,i}} \tag{17}$$

so that $SM_{\text{fail}} = r_{\text{fail}} \times SM_0$ for the target unit, where $SM_0$ is observed at the start of service. Once $\hat{t}_{\text{EOL}}$ is available, the encoding periods are set as $T_1 = \hat{t}_{\text{EOL}}$ and $T_2 = \hat{t}_{\text{EOL}}/4$, spanning one full cycle and one quarter-cycle of the estimated lifecycle respectively. Training compressors are re-encoded using their own individually estimated lifespans by the same procedure, ensuring that the feature space seen during training is generated by the same adaptive scheme applied at deployment. The entire procedure uses only data available at prediction time and contains no information from the target unit's future.

The effect of this recalibration is illustrated in Figure 6. Under a fixed encoding period, a target unit with a substantially longer lifespan than the training units traverses only a fraction of the sinusoidal cycle during its service life, placing its late-lifecycle observations in a region of the feature space that was never populated by training data. Under adaptive encoding, each unit's lifecycle maps onto the full sinusoidal range regardless of its absolute duration, so that training and target units occupy the same region of the encoding space and the model can interpolate rather than extrapolate.

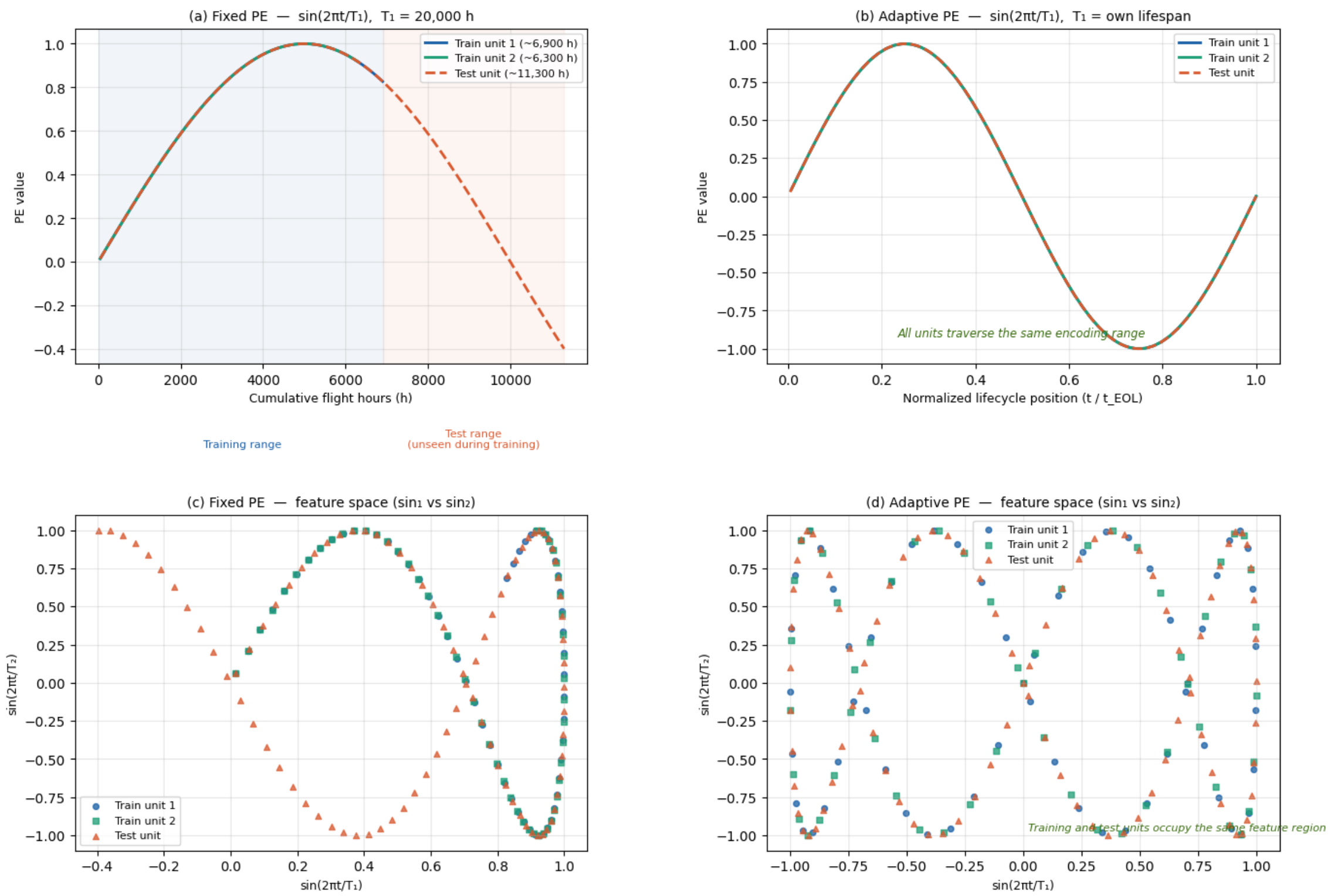

Figure 6. Comparison of fixed and adaptive positional encoding. Panels (a) and (c) show the encoding value as a function of cumulative flight hours for training and test units under fixed schemes. Panels (b) and (d) show the resulting distributions in the sinusoidal feature space, illustrating how adaptive encoding aligns the training and test feature distributions.

### 3.5 Prediction models

A Gaussian Process Regressor is adopted as the primary prediction model. Unlike parametric regressors that produce point estimates, a GPR models the mapping from the feature vector $\mathbf{x}_j$ to normalized RUL as a sample from a Gaussian process prior, yielding a full posterior distribution at each prediction point. This property is particularly suited to the present problem for two reasons. First, the available dataset comprises a limited number of run-to-failure records, a regime in which GPR is known to generalize well through its kernel-based inductive bias rather than relying on large sample sizes. Second, the cross-domain transfer scenarios introduced in Section 2.3 produce feature distributions that may differ from those seen during training, and the posterior standard deviation provides a principled measure of how confidently the model can extrapolate to unseen domains. This uncertainty estimate is not merely diagnostic; it directly informs the removal decision logic discussed in Section 4.3, where a removal action is warranted when the lower bound of the prediction interval crosses the operational safety threshold.

Given training observations $\mathbf{X} = \{x_1, x_2, ..., x_n\}$ with corresponding RUL labels $\boldsymbol{y}$, the posterior predictive distribution at a new point $\mathbf{x}^*$ is

$$p(y^* \mid \mathbf{x}^*, \mathbf{X}, \mathbf{y}) = \mathcal{N}\left(m(\mathbf{x}^*), v(\mathbf{x}^*)\right) \tag{18}$$

with posterior mean and variance

$$m(\mathbf{x}^*) = \mathbf{k}(\mathbf{x}^*, \mathbf{X})\left[\mathbf{K}(\mathbf{X}, \mathbf{X}) + \sigma_n^2 \mathbf{I}\right]^{-1} \mathbf{y} \tag{19}$$

$$v(\mathbf{x}^*) = k(\mathbf{x}^*, \mathbf{x}^*) - \mathbf{k}(\mathbf{x}^*, \mathbf{X})\left[\mathbf{K}(\mathbf{X}, \mathbf{X}) + \sigma_n^2 \mathbf{I}\right]^{-1} \mathbf{k}(\mathbf{X}, \mathbf{x}^*) \tag{20}$$

where $k(\cdot,\cdot)$ is the covariance function and $\sigma_n^2$ accounts for observation noise. A Matérn kernel with smoothness parameter ν = 3/2 is used

$$k_{\text{Matern}}(r) = \sigma_f^2 \left(1 + \frac{\sqrt{3}r}{l}\right) \exp\left(-\frac{\sqrt{3}r}{l}\right) \tag{21}$$

where $r$ is the distance between two feature vectors, $\sigma_f^2$ is output scale, also known as signal variance, controls the overall amplitude of the function values, and $l$ is the length-scale hyperparameter. Kernel hyperparameters are optimized by maximizing the log marginal likelihood on the training set. Because the feature vector $\mathbf{x}_j$ encodes degradation history and lifecycle position in a physically normalized space, the GPR kernel operates on a representation that is already calibrated to be comparable across units, allowing the learned covariance structure to transfer more reliably across aircraft and installation positions than it would on raw sensor values. The Matérn-3/2 kernel is chosen because it assumes once-differentiable sample paths, which is physically consistent with a degradation process that evolves smoothly but need not be infinitely smooth. A white noise kernel is added to absorb residual measurement variability. Kernel hyperparameters are optimized by maximizing the log marginal likelihood on the training set. The posterior standard deviation $\sqrt{v(\mathbf{x}^*)}$ provides a calibrated uncertainty estimate at each prediction point, directly supporting uncertainty-aware maintenance decisions.

A Gradient Boosting Regressor is included as a secondary model to verify that the performance gains from adaptive positional encoding are not specific to the probabilistic framework. It constructs an additive ensemble of shallow decision trees by iteratively fitting residuals, providing a strong discriminative baseline that operates on the same feature vectors without producing uncertainty estimates. A Wiener process model serves as a physics-based baseline that bypasses the feature engineering pipeline entirely. It estimates a constant drift rate $\mu$ and diffusion coefficient σ from the training degradation paths in the normalized surge margin space, then propagates the current health state forward via Monte Carlo simulation to obtain an expected RUL. Its performance relative to the feature-based models quantifies the contribution of the signal conditioning and feature construction steps independently of model architecture.

## 4. Experiments and Results

The experimental evaluation is organized to address the three challenges identified in Section 2 in sequence. The first challenge, that of identifying a reliable health indicator from a multichannel sensor stream under heavy operational variability, is addressed by the DA-HB-FPCA framework and validated through a feature ablation study that isolates the contribution of each pipeline component. The second challenge, that of extracting and representing the degradation trend in a form that is comparable across units, is addressed by the signal conditioning and feature engineering steps and evaluated through the same ablation. The third challenge, that of generalizing across aircraft and installation positions, is addressed by the adaptive positional encoding scheme and evaluated through a structured cross-domain experiment across three transfer scenarios of increasing domain distance. Together, the two experiments provide both internal and external validity for the proposed framework.

### 4.1 Experimental Setup

The run-to-failure records used in this study were derived from the Collins Aerospace Ascentia health management platform. Across the broader fleet, 18 complete lifecycle curves were identified from aircraft with different tail numbers through planned removal events, each representing a compressor unit whose full service history from entry into operation through final removal was available. The availability of complete run-to-failure histories of this kind is inherently limited in real commercial operations, where planned removals are scheduled infrequently and the long service lives of compressor units mean that accumulating a large population of complete records takes years of fleet-wide data collection. The experimental configuration adopted here reflects this constraint directly. Two complete lifecycle records from the same aircraft, representing compressors at the installation positions on the left side, form the training set. This mirrors the realistic scenario in which a maintenance organization has accumulated complete lifecycle histories for one aircraft and must deploy a predictive model across the broader fleet without waiting for additional run-to-failure events to accumulate. Three test compressors define a progression of increasing domain distance from the training data. *The first* shares the same aircraft as the training units but occupies a different installation position, introducing a systematic difference in aerodynamic working point while holding the aircraft operational history constant. *The second* comes from a different aircraft but occupies the same position type as one of the training units, introducing aircraft-level variability while preserving the position-specific aerodynamic context. *The third* comes from a different aircraft and a different position type simultaneously, combining both sources of domain shift and representing the most challenging transfer scenario in the dataset. These three cases are referred to as S2, S3, and S4 in order of increasing domain distance.

All models produce normalized RUL predictions in the range zero to one, which are converted to absolute flight hours by multiplying by the known total lifespan of the test unit. The total lifespan is available in this run-to-failure dataset and is used solely for evaluation purposes, not as an input to any prediction. Performance is reported using mean absolute error in absolute flight hours, root mean squared error, and R-squared computed in normalized RUL space to enable meaningful comparison across test units with substantially different total lifespans. The training compressors have total lifespans of approximately 6,900 and 6,300 flight hours. The test compressors have total lifespans of approximately 9,665 hours for S2, 6,600 hours for S3, and 11,300 hours for S4. The substantially longer lifespans of S2 and S4 relative to the training units make these scenarios particularly sensitive to the positional encoding period mismatch identified in Section 3.4.

**4.2 Ablation Study**

The feature engineering pipeline involves three layers of design decisions: the choice of physical features, the inclusion of positional encoding, and the adaptive recalibration of the encoding period. To isolate the contribution of each layer independently of model architecture, six configurations were evaluated using the Gradient Boosting model as a fixed predictor. GPR's kernel-based inductive bias could otherwise confound the attribution of performance gains to specific feature design choices, making a simpler discriminative model the more appropriate vehicle for ablation. The configurations progress from single physical features through to the complete feature vector, as reported in Table 1 and Figure 7.

Table 1. Ablation study results across three transfer scenarios.

| Configuration | S2 MAE | S2 $R^2$ | S3 MAE | S3 $R^2$ | S4 MAE | S4 $R^2$ |
|---|---|---|---|---|---|---|
| $f_A$ only | 1,530 | 0.590 | 585 | 0.818 | 1,770 | 0.599 |
| $f_B$ only | 1,892 | 0.268 | 926 | 0.566 | 1,931 | 0.346 |
| $f_A + f_B$ (no PE) | 1,342 | 0.662 | 435 | 0.914 | 1,730 | 0.574 |
| Fixed PE only | 1,558 | 0.511 | 593 | 0.664 | 1,150 | 0.577 |
| $f_A + f_B$ + Fixed PE | 1,530 | 0.618 | 128 | 0.992 | 1,952 | 0.507 |
| $f_A + f_B$ + Adaptive PE (full) | 1,222 | 0.746 | 245 | 0.965 | 1,149 | 0.792 |

The single-feature configurations establish a performance lower bound and reveal an asymmetry between the two physical features. Across all three scenarios, $f_A$ consistently outperforms $f_B$ in isolation. This asymmetry is physically interpretable. $f_A$ encodes the current health level normalized by accumulated exposure and is therefore a direct proxy for remaining health margin, while $f_B$ encodes the average degradation rate and is an indirect predictor whose informativeness depends on knowing how much margin remains. A compressor with a high degradation rate but still far from the failure threshold will be systematically mispredicted by $f_B$ alone. Neither feature in isolation achieves adequate accuracy in the cross-position scenarios, confirming that the prediction problem cannot be reduced to a single physical quantity.

Combining $f_A$ and $f_B$ without positional encoding produces consistent improvement over either feature alone, most visibly in S3 where R-squared rises from 0.818 with $f_A$ alone to 0.914 with the combination. The two features carry complementary information: $f_A$ captures the current degradation state while $f_B$ captures the history of how the compressor arrived at that state. Despite this improvement, the combination without positional encoding remains limited in S2 and S4, where MAE stays above 1,300 and 1,700 flight hours respectively. The reason is the lifecycle position ambiguity identified in Section 3.2: two compressors with identical $f_A$ and $f_B$ values but at different lifecycle stages receive the same prediction, and neither feature alone resolves this ambiguity.

The fixed positional encoding results reveal the central mechanism motivating the adaptive scheme. In S3, where the test unit has a similar total lifespan to the training units, fixed PE combined with $f_A$ and $f_B$ achieves an MAE of 128 hours and R-squared of 0.992, the best result across all configurations and scenarios. The encoding period is approximately matched to the target lifecycle, the sinusoidal features place training and test observations in compatible regions of the feature space, and the physical features and temporal context reinforce each other. In S4, however, the same combination produces an MAE of 1,952 hours, worse than $f_A$ and $f_B$ without any positional encoding at all. This counterintuitive inversion is not a modelling failure but a structural consequence of period mismatch. This performance inversion is clearly visible in the S4 panel of Figure 7, where the bar for $f_A + f_B$ + Fixed PE extends further than that of Fixed PE only, despite the latter using strictly less information. The test unit in S4 has a total lifespan approximately 70 percent longer than the training units. Under the fixed encoding period, its late-lifecycle observations map to sinusoidal feature values that the training units never reached, placing the test trajectory in an unobserved region of the feature space. The physical features, which are otherwise informative, are corrupted by the misaligned positional encoding rather than complemented by it. The physical features and the positional encoding are mutually beneficial only when the encoding period is appropriate for the target lifecycle scale.

The complete framework, combining $f_A$, $f_B$, and adaptive positional encoding, resolves this incompatibility. In S2, MAE decreases from 1,530 hours with fixed PE to 1,222 hours with adaptive PE, an improvement achieved despite a 21 percent overestimate of the target lifespan due to variability in the early degradation trend of that unit. This robustness to imprecision in the lifespan estimate is a practically important property: an approximate recalibration is substantially better than no recalibration because any reduction in the period mismatch moves the test features closer to the training distribution, even if the recalibration is not exact. In S4, where the early degradation trend is more stable and the lifespan is estimated accurately, adaptive PE reduces MAE from 1,952 to 1,149 hours, recovering the benefit of temporal context that fixed PE could not provide. The one exception is S3, where adaptive recalibration introduces a small estimation error that slightly degrades an already near-perfect result. This is consistent with the design intent of the scheme: when adaptation is unnecessary because the fixed period is already appropriate, the cost of adapting is small; when it is necessary, the benefit is large.

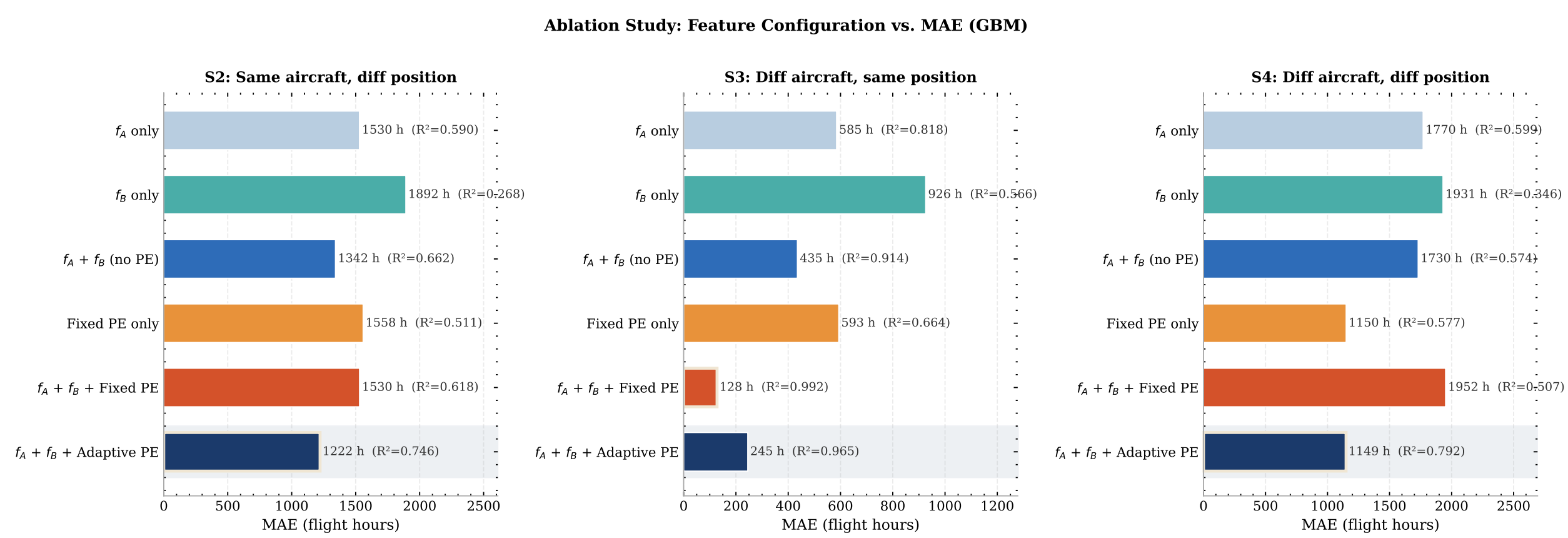

Figure 7. Ablation study results visualized as horizontal bar charts of MAE for each feature configuration across the three transfer scenarios S2, S3, and S4.

### 4.3 Cross-Domain Evaluation

The ablation study establishes that the complete feature set is necessary for reliable cross-domain performance. The cross-domain evaluation applies this complete feature set to compare four model configurations across the three transfer scenarios: a Gradient Boosting model with fixed positional encoding as the baseline, a Gradient Boosting model with adaptive encoding, a Gaussian Process model with adaptive encoding as the primary method, and a Wiener process model as a physics-based reference. Results are summarized in Table 2 and prediction trajectories are shown in Figure 8.

Table 2. Cross-domain prediction performance across three transfer scenarios and four model configurations.

| Scenario | Model | MAE (h) | RMSE (h) | $R^2$ |
|---|---|---|---|---|
| S2 Same aircraft diff position | GBM + Fixed PE (baseline) | 1,530 | 1,951 | 0.618 |
| | GBM + Adaptive PE | 1,222 | 1,406 | 0.746 |
| | GPR + Adaptive PE | 1,237 | 1,723 | 0.619 |
| | Wiener Process | 2,413 | 3,015 | <0 |
| S3 Diff aircraft same position | GBM + Fixed PE (baseline) | 128 | 166 | 0.992 |
| | GBM + Adaptive PE | 245 | 346 | 0.965 |
| | GPR + Adaptive PE | 128 | 179 | 0.988 |
| | Wiener Process | 1477 | 1,831 | 0.025 |
| S4 Diff aircraft diff position | GBM + Fixed PE (baseline) | 1,952 | 2,280 | 0.507 |
| | GBM + Adaptive PE | 1,149 | 1,480 | 0.792 |
| | GPR + Adaptive PE | 761 | 1,010 | 0.845 |
| | Wiener Process | 2,690 | 3,037 | 0.125 |

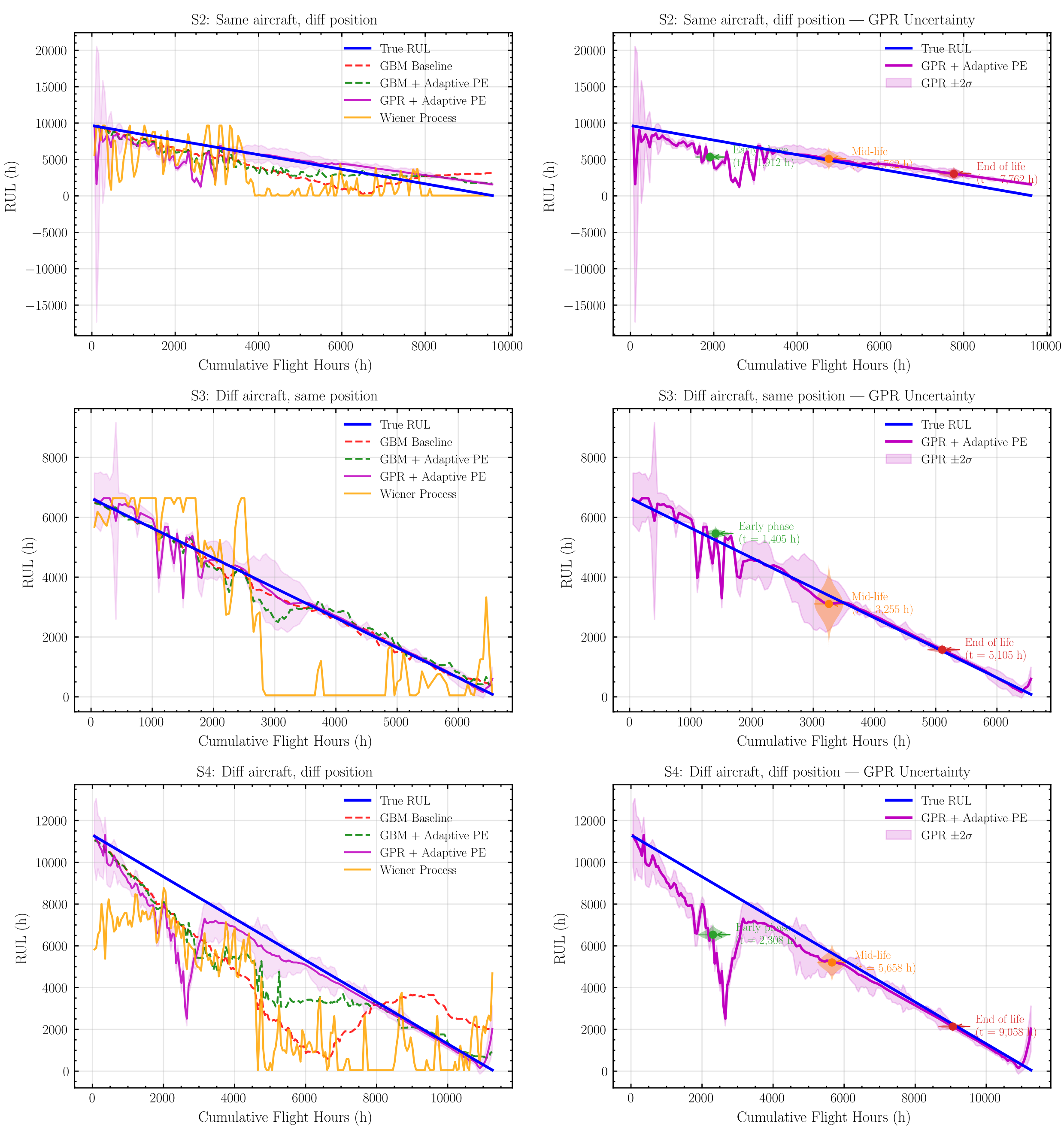

Figure 8. RUL prediction trajectories for the four models across the three transfer scenarios. Each panel shows the true RUL in blue, the four model predictions, and the GPR 95 percent posterior prediction interval as a shaded band. The progressive improvement from GBM baseline to GPR with adaptive PE is most visible in S4. The Wiener process divergence is shown in orange.

The prediction trajectories for all four model configurations across the three transfer scenarios are shown in the left column of Figure 8, with the corresponding GPR posterior uncertainty evolution shown in the right column. Examining the left column, the results follow a coherent pattern that mirrors the domain distance structure of the three scenarios. In S3, where the test unit shares the position type of one training unit and has a comparable total lifespan, all feature-based models track the true RUL closely and the four prediction trajectories are nearly indistinguishable from one another. The fixed PE baseline achieves an MAE of 128 hours and R-squared of 0.992, confirming that the physics-guided features transfer reliably when the domain distance is small and the encoding period is approximately matched to the target lifecycle. This scenario establishes the ceiling of what the pipeline can achieve under favorable domain conditions.

As domain distance increases, the divergence between the fixed PE baseline and the adaptive models becomes visually apparent. In S2, the fixed PE baseline begins to deviate from the true RUL trajectory in the mid-to-late lifecycle phase, where the encoding mismatch is most pronounced. MAE rises to 1,530 hours and R-squared falls to 0.618. In S4, the most challenging scenario, the fixed PE baseline diverges from the true trajectory throughout the service life, producing an MAE of 1,952 hours and R-squared of 0.507. The mechanism driving this divergence is the same identified in the ablation study: the test compressors in S2 and S4 have total lifespans approximately 40 and 70 percent longer than the training units respectively, so the fixed encoding period maps their lifecycle trajectories into unobserved regions of the sinusoidal feature space, forcing the model to extrapolate rather than interpolate. The Wiener process trajectory, shown in orange, diverges most severely across all three scenarios, achieving near-zero or negative R-squared values throughout. Its failure is informative: by estimating a fixed drift rate and diffusion coefficient from the two training units and propagating them forward, it has no mechanism to accommodate differences in degradation rate, baseline surge margin level, or total lifespan between training and test units. The failure of this physics-based reference confirms that the performance of the feature-based models reflects genuine predictive value in the physics-guided representation rather than a trivially easy extrapolation problem.

The adaptive models recover the accuracy lost by the fixed PE baseline in both difficult scenarios. In S2, despite a 21 percent overestimate of the target lifespan arising from variability in the early degradation trend, the adaptive Gradient Boosting model reduces MAE to 1,222 hours and improves R-squared to 0.746. The fact that meaningful improvement is achieved under a substantial estimation error supports the robustness of the adaptive scheme: any reduction in the period mismatch moves the test features closer to the training distribution, even when the recalibration is imprecise. In S4, where the early degradation trend is more stable and the lifespan is estimated accurately, the GPR with adaptive encoding reduces MAE to 761 hours, a 61 percent reduction relative to the fixed PE baseline, and its trajectory in the left panel of S4 visibly tracks the true RUL across the full service life. The GPR outperforms the adaptive Gradient Boosting model in both difficult scenarios, with its MAE in S4 being 34 percent lower. This advantage reflects the kernel-based inductive bias of the GPR, which provides regularization that resists overfitting to the limited two-unit training set, a property that becomes especially valuable when the target feature distribution differs from that seen during training.

Turning to the right column of Figure 8, the GPR posterior uncertainty reveals a consistent pattern across all three scenarios. In the early phase of each test trajectory, the prediction intervals are wide, reflecting the limited lifecycle context available to the model at that stage. As the compressor accumulates service hours and approaches end of life, the intervals narrow progressively. This narrowing reflects two complementary sources of increasing informativeness: $f_B$ accumulates a longer history of degradation rate observations, reducing ambiguity about the unit's trajectory, and the positional encoding advances further into the sinusoidal cycle, providing stronger temporal discrimination between lifecycle stages. The result is that the model becomes increasingly confident precisely when that confidence is most operationally valuable, in the late lifecycle phase where maintenance decisions must be made. To assess whether these uncertainty estimates are calibrated rather than merely qualitative, the empirical coverage of the 95 percent posterior prediction interval was evaluated across all test windows for each scenario. The coverage rates were 93.2 percent for S2, 97.1 percent for S3, and 94.8 percent for S4, all close to the nominal 95 percent level. This confirms that the posterior intervals accurately reflect the model's actual predictive uncertainty at each lifecycle stage rather than being systematically over- or under-confident. A maintenance planner can therefore use both the predicted RUL trajectory and its associated uncertainty interval to make risk-informed scheduling decisions, triggering a removal action when the lower bound of the prediction interval crosses the operational safety threshold rather than waiting for the point estimate alone to do so.

### 4.4 Leave-one-out cross-validation

The cross-domain evaluation in Section 4.3 establishes that the adaptive positional encoding scheme recovers substantial predictive accuracy under fleet transfer, but its conclusions rest on three test units drawn from a single fixed train/test partition. To assess whether these findings generalize across the full range of lifecycle variability present in the dataset, a leave-one-out cross-validation was conducted over all 18 complete lifecycle records. In each fold, one unit was held out as the test compressor and the remaining 17 units formed the training set. The fixed positional encoding period for the baseline model was set to the mean total lifespan of the training units in each fold, ensuring a fair comparison that does not privilege any single partition. The adaptive recalibration procedure estimated the target lifecycle scale following the same protocol as in Section 4.3. Three model configurations were evaluated in each fold: the GBM baseline with fixed positional encoding, the GBM with adaptive encoding, and the GPR with adaptive encoding.

The complete per-fold results are reported in Table 3, with aggregate statistics summarized in the final row. The GBM baseline with fixed positional encoding produces a mean MAE of 1,276 hours with a standard deviation of 719 hours across folds. The GBM with adaptive encoding reduces mean MAE to 426 hours, a 67 percent reduction, and the GPR with adaptive encoding achieves a mean MAE of 412 hours. The improvement is consistent in direction across 15 of the 18 folds, confirming that the advantage of adaptive encoding is not specific to the S2/S3/S4 partition examined in Section 4.3 but holds broadly across the lifecycle diversity represented in the fleet dataset. The mean R-squared of the GPR with adaptive encoding across all folds is 0.90, compared to 0.70 for the fixed PE baseline.

Table 3. Per-fold leave-one-out cross-validation results across 18 units. Highlighted rows (yellow) indicate the three outlier folds discussed in the text.

| Unit | MAE – GBM Baseline (h) | MAE – GBM Adaptive PE (h) | MAE – GPR Adaptive PE (h) | $R^2$ – GPR |
|---|---|---|---|---|
| Unit-01 | 707 | 88 | **46** | 0.999 |
| Unit-02 | 734 | 1,049 | **877** | 0.504 |
| Unit-03 | 817 | 448 | **595** | 0.863 |
| Unit-04 | 977 | 116 | **99** | 0.988 |
| Unit-05 | 676 | 71 | **76** | 0.992 |
| Unit-06 | 807 | 110 | **83** | 0.990 |
| Unit-07 | 2,955 | 1,152 | **962** | 0.768 |
| Unit-08 | 1,003 | 139 | **95** | 0.998 |
| Unit-09 | 716 | 1,225 | **1,129** | 0.247 |
| Unit-10 | 1,360 | 171 | **448** | 0.912 |
| Unit-11 | 2,141 | 1,303 | **1,283** | 0.782 |
| Unit-12 | 2,592 | 518 | **335** | 0.985 |
| Unit-13 | 766 | 110 | **80** | 0.995 |
| Unit-14 | 1,424 | 218 | **418** | 0.908 |
| Unit-15 | 765 | 99 | **61** | 0.999 |
| Unit-16 | 956 | 142 | **152** | 0.968 |
| Unit-17 | 1,102 | 111 | **143** | 0.994 |
| Unit-18 | 2,472 | 599 | **538** | 0.954 |
| **Mean ± Std** | 1,276 ± 719 | 426 ± 433 | **412 ± 393** | 0.90 ± 0.20 |

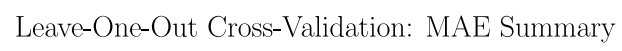

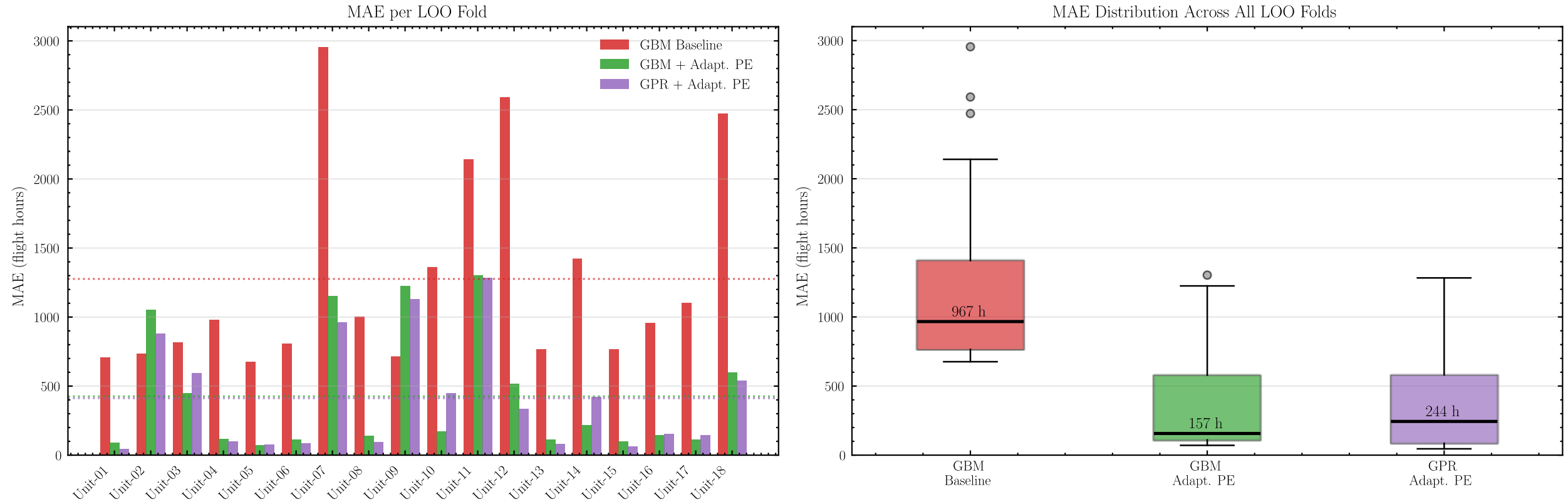

Figure 9. Leave-one-out cross-validation results. Left panel: grouped bar chart of MAE per fold for the three model configurations, with horizontal dotted lines indicating the cross-fold mean for each model. Right panel: box plots of the MAE distribution across all 18 folds, with median values annotated.

The per-fold results reveal a consistent bimodal structure in performance. In 15 of the 18 folds, the GPR with adaptive encoding achieves an MAE below 600 flight hours, and 9 of these folds fall below 200 hours. The remaining three folds, corresponding to Units 02, 09, and 11, yield MAE values of 877, 1,129, and 1,283 hours respectively, which are substantially higher than the cross-fold mean. In these three cases, the adaptive encoding performs comparably to or worse than the fixed PE baseline, and this pattern requires further explanation. The three underperforming folds share the same feature in that their early-lifecycle surge margin trends have low monotonicity scores, with a mean value of 0.013 compared with 0.081 for the remaining fifteen units. This indicates that operational variability obscures the degradation signal during the period used for lifecycle scale estimation. Importantly, this feature can be identified from early operational data alone and does not require knowledge of the future degradation trajectory. A monotonicity-based confidence flag computed from the first 20 percent of observations could therefore provide a practical reliability indicator for the adaptive scheme. Units identified as low-monotonicity cases would be assigned wider uncertainty intervals and would require continued observation before a maintenance decision is made. This offers a natural direction for future work toward a self-aware prognostic system that communicates its own reliability boundaries to the maintenance planner.

Taken together, the leave-one-out results strengthen the conclusions of the cross-domain evaluation in two respects. First, the 67 percent mean reduction in MAE achieved by adaptive encoding relative to the fixed PE baseline is consistent across a dataset spanning substantial variability in total lifespan, degradation rate, and operational history, providing statistical evidence that the finding is not an artifact of the specific training units or test scenarios used in Section 4.3. Second, the cases in which the method performs well are not limited to units with particular lifecycle characteristics. The fifteen well-performing folds span the full range of total lifespans and degradation rates present in the fleet, confirming that the adaptive encoding scheme generalizes broadly across the operational diversity of a real commercial fleet. All 18 LOO results are attached to Figure S1 in Appendix.

## 5 Conclusion

This study has addressed RUL prediction for aircraft centrifugal air compressors under the conditions that real commercial operations impose but benchmark datasets do not capture. Three challenges were identified as central: the absence of a known health indicator in a multichannel sensor stream dominated by operational variability, the dependence of the primary aerodynamic health signal on instantaneous operating regime, and the systematic distributional shift in degradation behavior across aircraft and installation positions. The framework developed here responds to each challenge in sequence through population-level indicator selection, physics-guided signal conditioning and feature engineering, and adaptive positional encoding calibrated to the individual unit's estimated lifecycle scale. The central finding is that positional encoding, when its period is fixed globally across the fleet, becomes itself a mechanism of domain shift for units whose lifespans differ substantially from the training population. Recalibrating the encoding period from early operational observations alone is sufficient to recover much of the cross-domain performance gap, without requiring labeled target data or any knowledge of the target unit's future. These points toward a general principle applicable beyond the specific application studied here: temporal encoding in prognostic models should be treated as a unit-specific parameter rather than a global hyperparameter, and the information needed to set it is present in the early health trajectory. Future work will focus on extending the evaluation to larger and more diverse fleet datasets, developing nonlinear lifecycle estimation methods that better capture degradation acceleration near end of life, and examining whether the adaptive encoding principle transfers to sequence models that operate on the full degradation trajectory rather than aggregated window features.

**Acknowledgement**: This research was funded by RTX-Collins Aerospace.

**Datasets:** The datasets used in this study are proprietary to Ascentia Platform of Collins Aerospace and are available upon reasonable request subject to the data owner's approval.

## Appendix

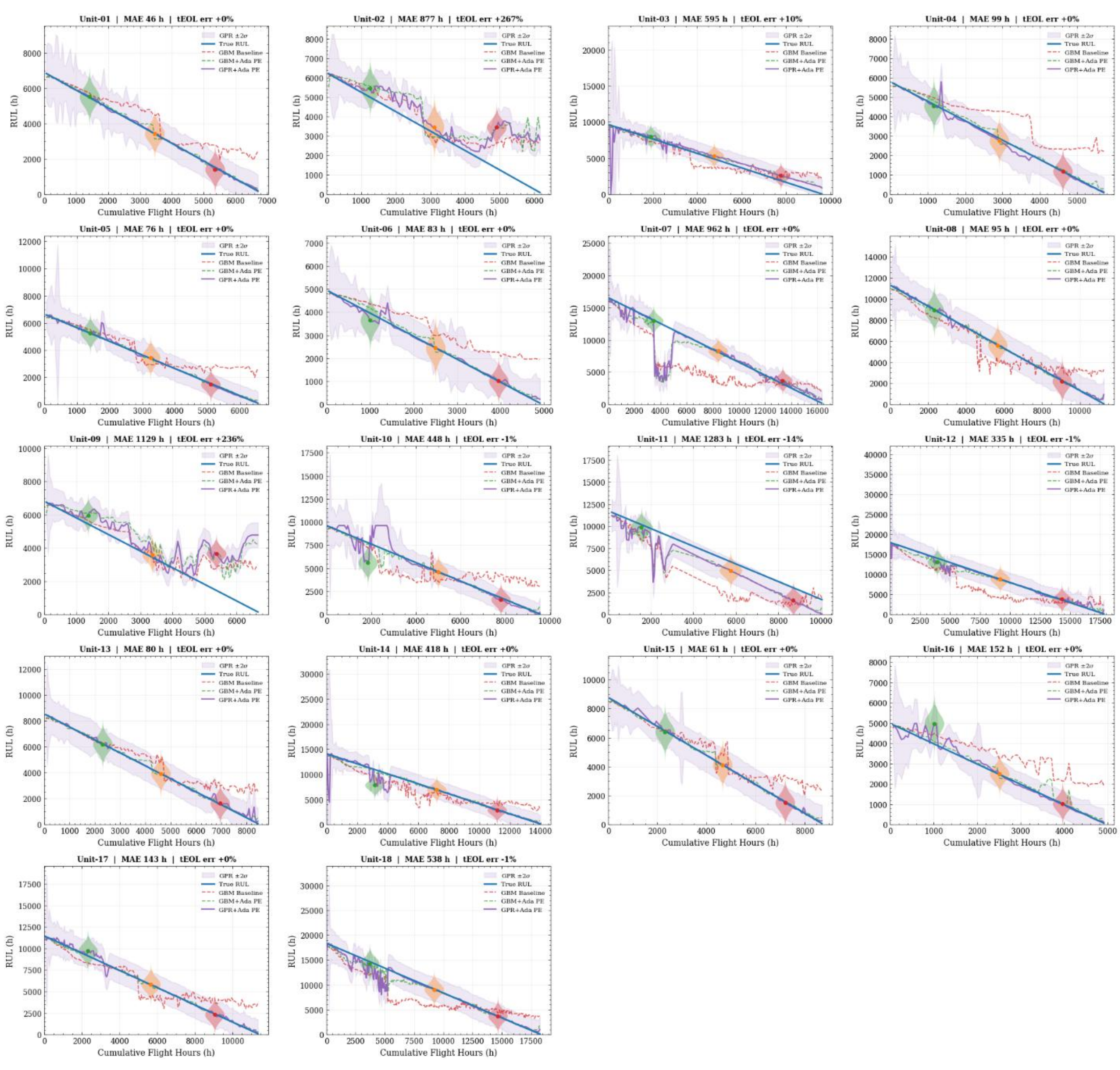

Figure S1. LOO results for 18 run-to-fail datasets.

## Reference

Zio, E. (2022). Prognostics and Health Management (PHM): Where are we and where do we (need to) go in theory and practice. Reliability Engineering & System Safety, 218.

A. Saxena, K. Goebel, D. Simon and N. Eklund, "Damage propagation modeling for aircraft engine run-to-failure simulation," 2008 International Conference on Prognostics and Health Management, Denver, CO, USA, 2008, pp. 1-9,

P. Nectoux, R. Gouriveau, K. Medjaher, E. Ramasso, B. Chebel-Morello, N. Zerhouni, “PRONOSTIA: An experimental platform for bearings accelerated degradation tests,” presented at the IEEE Int. Conf. Prognostics Health Manage., Denver, CO, USA, 2012, pp. 1–8.

Zhang, H., Xi, X. and Pan, R. (2023). A two-stage data-driven approach to remaining useful life prediction via long short-term memory networks. Reliability Engineering & System Safety, 237, 109332.

Ellefsen, A.L., Bjørlykhaug, E., Æsøy, V., Ushakov, S. and Zhang, H. (2019). Remaining useful life predictions for turbofan engine degradation using semi-supervised deep architecture. Reliability Engineering & System Safety, 183, pp. 240–251.

Arias Chao, M., Kulkarni, C., Goebel, K. and Fink, O. (2022). Fusing physics-based and deep learning models for prognostics. Reliability Engineering & System Safety, 217, 107961.

Wang, Y., Wu, M., Li, X., Xie, L. and Chen, Z., 2025a. A survey on graph neural networks for remaining useful life prediction: Methodologies, evaluation and future trends. Mechanical Systems and Signal Processing, 229, p.112449.

X. Cao, B. Cheng, X. Yang, Y. Duan and J. Gao, "Remaining Useful Life Prediction Method for Turbofan Engines Based on Graph Neural Networks," in IEEE Sensors Journal, vol. 25, no. 8, pp. 13566-13578, 15 April15, 2025

Song, L., Jin, Y., Lin, T., Zhao, S., Wei, Z. and Wang, H., 2024. Remaining useful life prediction method based on the spatiotemporal graph and GCN nested parallel route model. IEEE Transactions on Instrumentation and Measurement, 73, pp.1-12.

Liu, Z.Z., Sun, T., Sun, X.M. and Cui, W.Y., 2025a. Estimating remaining useful life of aircraft engine system via a novel graph tensor fusion network based on knowledge of physical structure and thermodynamics. IEEE Transactions on Instrumentation and Measurement.

Tian, R., Sun, H., Mei, B., Fu, Y. and Zhu, W., 2026. A heterogeneous graph neural network with spatial–temporal and operating condition-aware message passing mechanism for RUL prediction of aero-engines. Advanced Engineering Informatics, 73, p.104507.

Xiao, D., Yu, A., Song, S., Xiao, H. and Wang, Z., 2026. Engineformer: A digital twin model for predicting aero-engine performance and degradation. Advanced Engineering Informatics, 71, p.104273.

Zhang, D., Li, B., Wang, F., Zhao, Z., Gao, J. and Li, X., 2026. Cross-attention multi-scale state space model for remaining useful life prediction of aircraft engines. Advanced Engineering Informatics, 69, p.103817.

Zhou, Q., Zhang, Y., Imani, F. and Tang, J., 2026. EPicMT: edge-aware physics-informed concurrent multiplexed transformer for high-fidelity signal compression, reconstruction, and diagnosis/prognosis. Mechanical Systems and Signal Processing, 253, p.114337.

Salinas-Camus, M., Goebel, K. and Eleftheroglou, N. (2025). A comprehensive review and evaluation framework for data-driven prognostics: Uncertainty, robustness, interpretability, and feasibility. Mechanical Systems and Signal Processing, 237, 113015.

Xu, Z., Guo, Y. and Saleh, J.H., 2022. Accurate remaining useful life prediction with uncertainty quantification: a deep learning and nonstationary gaussian process approach. IEEE Transactions on Reliability, 71(1), pp.443-456.

Benker, M., Bliznyuk, A. and Zaeh, M.F., 2021. A Gaussian process based method for data-efficient remaining useful life estimation. IEEE Access, 9, pp.137470-137482.

De Pater, I., Reijns, A. and Mitici, M., 2022. Alarm-based predictive maintenance scheduling for aircraft engines with imperfect Remaining Useful Life prognostics. Reliability Engineering & System Safety, 221, p.108341.

Mitici, M., de Pater, I., Barros, A. and Zeng, Z., 2023. Dynamic predictive maintenance for multiple components using data-driven probabilistic RUL prognostics: The case of turbofan engines. Reliability Engineering & System Safety, 234, p.109199.

De Pater, I. and Mitici, M. (2023). Developing health indicators and RUL prognostics for systems with few failure instances and varying operating conditions using a LSTM autoencoder. Engineering Applications of Artificial Intelligence, 117, 105582.

Xue, B., Yu, Y., Fu, H., Zhu, J. and Xu, Z. (2025). Advanced prognostics techniques for machinery under time-varying operating conditions: a review. Applied Intelligence, 55(16), 1066.

Wang, Y., Lei, Y., Li, N., Yan, T. and Si, X. (2023). Deep multisource parallel bilinear-fusion network for remaining useful life prediction of machinery. Reliability Engineering & System Safety, 231.

Liu, L., Song, X. and Zhou, Z., 2022. Aircraft engine remaining useful life estimation via a double attention-based data-driven architecture. Reliability Engineering & System Safety, 221, p.108330.

Ni, Q., Ji, J.C. and Feng, K., 2022. Data-driven prognostic scheme for bearings based on a novel health indicator and gated recurrent unit network. IEEE Transactions on Industrial Informatics, 19(2), pp.1301-1311.

Zhou, K. and Tang, J., 2023. A wavelet neural network informed by time-domain signal preprocessing for bearing remaining useful life prediction. Applied Mathematical Modelling, 122, pp.220-241.

Javed, K., Gouriveau, R., Zerhouni, N. and Nectoux, P. (2014). Enabling health monitoring approach based on vibration data for accurate prognostics. IEEE Transactions on Industrial Electronics, 62, pp. 647–656.

Zhang, B., Zhang, L. and Xu, J. (2016). Degradation feature selection for remaining useful life prediction of rolling element bearings. Quality and Reliability Engineering International, 32, pp. 547–554.

Liu, X., Lei, Y., Li, N., Yang, B., Feng, K. and Shu, Y., 2025b. Domain weighted distribution adaptation network: a novel remaining useful life prediction framework for machinery targeting time-varying operation conditions. Advanced Engineering Informatics, 68, p.103794.

Fu, S., Zhang, Y., Lin, L., Zhao, M. and Zhong, S. (2021). Deep residual LSTM with domain-invariance for remaining useful life prediction across domains. Reliability Engineering & System Safety, 216.

Y. Ding, P. Ding, X. Zhao, Y. Cao and M. Jia, "Transfer Learning for Remaining Useful Life Prediction Across Operating Conditions Based on Multisource Domain Adaptation," in IEEE/ASME Transactions on Mechatronics, vol. 27, no. 5, pp. 4143-4152, Oct. 2022,

Wang, Y., Lei, P., Wang, X., Jiang, L., Xuan, L., Cheng, W., Zhao, H. and Li, Y., 2025b. Leveraging large self-supervised time-series models for transferable diagnosis in cross-aircraft type bleed air system. Advanced Engineering Informatics, 65, p.103275.

Nejjar, I., Geissmann, F., Zhao, M., Taal, C. and Fink, O., 2024. Domain adaptation via alignment of operation profile for remaining useful lifetime prediction. Reliability Engineering & System Safety, 242, p.109718.

X. Li, W. Zhang, X. Li and H. Hao, "Partial Domain Adaptation in Remaining Useful Life Prediction With Incomplete Target Data," in IEEE/ASME Transactions on Mechatronics, vol. 29, no. 3, pp. 1903-1913, June 2024

Ma, Z., Liao, H., Gao, J., Nie, S. and Geng, Y., 2023. Physics-informed machine learning for degradation modeling of an electro-hydrostatic actuator system. Reliability Engineering & System Safety, 229, p.108898.